\documentclass[preprint,superscriptaddress,amsmath,amssymb,aps,notitlepage,floatfix,longbibliography]{revtex4-2}
\usepackage{graphicx}
\usepackage{comment}
\usepackage{caption}
\usepackage{upgreek}
\usepackage{gensymb}
\usepackage{float}
\usepackage{bm}
\usepackage{xcolor}
\usepackage{physics}
\usepackage{multirow}
\usepackage{array}
\usepackage{makecell}
\usepackage{siunitx}
\usepackage[normalem]{ulem}
\usepackage[export]{adjustbox} 

\usepackage{xr}
\makeatletter
\newcommand*{\addFileDependency}[1]{
	\typeout{(#1)}
	\@addtofilelist{#1}
	\IfFileExists{#1}{}{\typeout{No file #1.}}
}
\makeatother

\newcommand*{\myexternaldocument}[1]{
	\externaldocument{#1}
	\addFileDependency{#1.tex}
	\addFileDependency{#1.aux}
}

\myexternaldocument{supps_071126}

\begin{document}

\title{Tellurium Metasurface Beam Splitter with Pulse Laser-Controlled Anisotropy}

\author{Takuto Hiraoka}
\affiliation{Department of Physics, The University of Tokyo, Tokyo 113-0033, Japan}

\author{Mizuho Matoba}
\affiliation{Institute for Photon Science and Technology, The University of Tokyo, Tokyo 113-0033, Japan}

\author{Yuta Kobayashi}
\affiliation{Department of Physics, The University of Tokyo, Tokyo 113-0033, Japan}

\author{Arata Mitsuzuka}
\affiliation{Department of Physics, The University of Tokyo, Tokyo 113-0033, Japan}

\author{Masashi Kawaguchi}
\affiliation{Department of Physics, The University of Tokyo, Tokyo 113-0033, Japan}

\author{Haruyuki Sakurai}
\affiliation{Institute for Photon Science and Technology, The University of Tokyo, Tokyo 113-0033, Japan}

\author{Kuniaki Konishi}
\affiliation{Institute for Photon Science and Technology, The University of Tokyo, Tokyo 113-0033, Japan}

\author{Masamitsu Hayashi}
\affiliation{Department of Physics, The University of Tokyo, Tokyo 113-0033, Japan}
\affiliation{Trans-Scale Quantum Science Institute, The University of Tokyo, Bunkyo, Tokyo 113-0033, Japan}

\date{\today}

\begin{abstract}
Laser-programmable optical anisotropy offers a new route to developing reconfigurable metasurfaces without conventional nanofabrication processes. Here, we demonstrate a lithography-free approach based on spatial control of the crystallographic $c$ axis orientation in tellurium (Te) using pulse laser irradiation. As a proof of concept, we demonstrate a Te metasurface beam splitter by laser-written optical-axis patterning and experimentally confirm that its optical response is in good agreement with theoretical predictions and numerical simulations. 
By directly programming the local optical anisotropy, this method enables a simple fabrication process while offering the possibility of rewriting and dynamically reconfiguring device functionality. These features make this approach a promising platform for non-resonant active metasurfaces and other reconfigurable flat-optics applications.
\end{abstract}

\maketitle
\clearpage
\section{Introduction}
As products such as smartphones and laptops become thinner and lighter, the demand for compact and lightweight optical components is increasing. Conventional optical elements, such as lenses and mirrors, are often bulky and heavy; therefore, compact and lightweight optical platforms are required as alternatives.
In this context, metasurfaces have attracted considerable attention as a new type of optical component. A metasurface consists of a two-dimensional array of subwavelength structures that can manipulate the phase, amplitude, and polarization of light\cite{yu2014nmat,genevet2017optica,kamali2018nanophoton}. By designing the geometry and arrangement of these structures, a wide range of optical functions, including beam steering\cite{cheng2015advoptmater,hu2015nanotechnol,yu2024scirep}, focusing\cite{hu2013advoptmater,chen2014advoptmater,khorasaninejad2016science,yang2024resphys}, optical vortex generation\cite{yu2011science,huang2012nanolett}, and holography\cite{choi2019advoptmater}, can be realized. 

Achieving these functions with high efficiency generally requires careful engineering of the phase and amplitude responses of individual subwavelength elements. In conventional metasurface design, this is typically achieved by numerical simulations of candidate structures, followed by patterning of the optimized structures using advanced nanofabrication techniques. This design-and-fabrication strategy has enabled high-performance metasurfaces based on engineered nanoscale structures that include high-aspect-ratio pillars\cite{khorasaninejad2016polarization,wang2021high,she2018large,park2019all}. Alternative fabrication routes have also been explored. For example, nanoimprint lithography has been widely explored as a scalable approach for fabricating flat optical elements by replicating nanoscale patterns in imprint resist layers\cite{einck2021scalable,gong2023simple,hoang2026300}. Because the pattern is replicated from a reusable mold, this approach is attractive for large-area, low-cost production. Beyond such mold-based replication strategies, resist-based approaches that directly exploit the optical functionality of the patterned material offer another route to simplified fabrication. Indeed, previous studies have shown that exposed resist itself can be used as the optical material and that diffractive flat optics can be fabricated directly from functional photoresists\cite{andren2020large,yamada2025optical}. 
Significant effort has therefore been directed toward simplifying the fabrication workflow to achieve low-cost production methods.

Functional reconfigurability is also an important theme in metasurface research. In conventional geometry-defined metasurfaces, the optical response is primarily determined by the shape and arrangement of the subwavelength structures, so the implemented function is typically defined at the fabrication stage. Reconfigurable metasurfaces have therefore been widely explored using externally controllable materials, including phase-change materials\cite{shalaginov2021ncomm,cai2022advphotonres}, liquid crystals\cite{komar2018acsphoton,he2018optlett}, and two-dimensional materials\cite{li2023nanolett}, in combination with patterned subwavelength structures. These studies highlight the broader possibility of defining metasurface functionality not only by fixed geometrical structures, but also by controllable material responses. 
\begin{figure}[htbp]
\centering
\includegraphics[width=0.95\columnwidth]{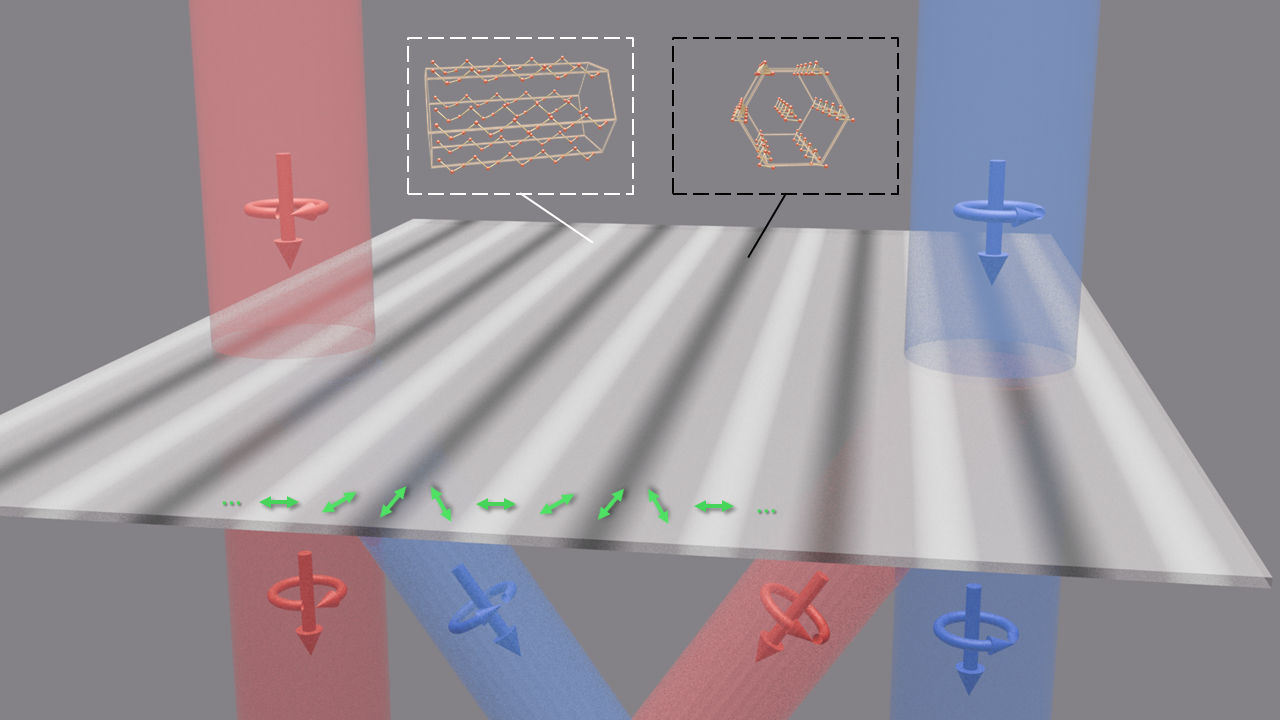}
\caption{\label{fig:beamsplitter}
A schematic illustration of the Te beam splitter. The $c$ axis of the Te film is spatially controlled by laser pulses. The greyscale contrast in the flat Te film and the corresponding green double-headed arrows represent spatial variation of the laser-written $c$ axis orientation, which enables the beam-splitting function. When the laser-written film is illuminated with left (right)-circularly polarized  light, part of the transmitted light is deflected due to the spatially varying $c$ axis orientation of Te. The deflected light is right (left)-circularly polarized. 
Inset: schematic illustration of Te crystal.  Te forms helical chains linked by covalent bonds along the $c$ axis, while adjacent chains are weakly bound by van der Waals interactions along the $a$ and $b$ axes. The white (black)-framed inset shows a Te crystal with the $c$ axis lying horizontally in the image plane (oriented along the viewing direction). 
} 
\end{figure}

Material responses therefore provide an additional design degree of freedom for engineering metasurface functionality. Laser writing offers a direct means of exploiting this capability by locally tailoring the optical properties of a material. Previous studies have demonstrated laser-written control of optical responses, including fabrication of birefringent patterns in glass for flat optical elements and manipulation of grain orientation and polarization-dependent optical properties in Ge$_2$Sb$_2$Te$_5$ films\cite{zhou2018broadband,meng2020tunable}.
While these studies established laser writing as a versatile tool for tailoring material responses, its application to wavefront control through laser-written anisotropy in intrinsically anisotropic thin films remains largely unexplored.

In this context, tellurium (Te) is an attractive material platform. Elemental Te is a narrow-band-gap semiconductor with a band gap of approximately $0.34 \ \si{\eV}$\cite{wang2025mid}. Te exhibits a high refractive index together with relatively low optical absorption in the near-infrared region, making it suitable for low-loss optical phase manipulation. In addition, Te exhibits strong uniaxial anisotropy in both the refractive index and the absorption coefficient in the visible to near infrared range, which is highly advantageous for polarization-dependent optical control\cite{guo2022nanoscale}. This anisotropy originates from its helical chainlike crystal structure and the resulting structural anisotropy\cite{momma2011japplcryst}. These optical properties make Te a promising platform for anisotropy-based optical control.

Moreover, it has been reported recently that the anisotropy axis of Te, namely the crystallographic $c$ axis, can be reoriented by irradiation with intense linearly polarized light pulses. Previous study has shown that exposure to $1030 \ \si{\nano \meter}$ linearly polarized pulse laser induces preferential alignment of the $c$ axis perpendicular to the incident polarization direction\cite{kobayashi2026nanolett}. Since the reorientation is induced locally by laser irradiation, the in-plane $c$ axis distribution can be spatially rewritten after film growth by scanning the laser spot. Interestingly, it has been demonstrated that the $c$ axis orientation can be overwritten by repeated irradiation, allowing reconfiguration of the anisotropy patterns, e.g. the grating of the metasurface\cite{kobayashi2026nanolett}. This light-induced and rewritable control of the anisotropy suggests a direct and straightforward route to metasurfaces whose optical functionality can be engineered and even reconfigured.

Here, we demonstrate the potential of laser-written Te as a metasurface platform by fabricating a beam deflector and evaluating its optical performance. We show that the metasurface can be formed by controlling the $c$ axis of the Te film using linearly polarized light pulses. The fabricated beam splitter, as illustrated in Fig.~\ref{fig:beamsplitter}, deflects light in the designed direction with the expected efficiency. These results establish Te as a promising material for metasurface-based optical elements that are both easy to fabricate and reconfigurable.

\section{Model}
Beam steering in metasurfaces originates from a spatially varying Pancharatnam-Berry (PB) phase\cite{li2017nonlinear}. In contrast to the propagation phase, which depends on the optical path length, the PB phase is determined by the in-plane orientation of the $c$ axis. Therefore, spatial modulation of the optic axis (i.e. the $c$ axis) in Te provides a simple route to imposing a phase gradient on light.

First, we consider a locally uniform birefringent element whose in-plane optic axis is rotated by an angle $\theta$ from the $x$ axis. For the transmission geometry, the Jones matrix is written as
\begin{align}
J(\theta)=R(-\theta)
\begin{bmatrix}
t_o & 0\\
0 & t_e
\end{bmatrix}
R(\theta),
\end{align}
where $t_o$ and $t_e$ are the complex amplitude transmission coefficients.
An analogous expression holds for the reflection geometry by replacing $t_o$ and $t_e$ with the complex amplitude reflection coefficients $r_o$ and $r_e$. The rotation matrix $R(\theta)$ is given by
\begin{align}
R(\theta)=
\begin{bmatrix}
\cos\theta & -\sin\theta\\
\sin\theta & \cos\theta
\end{bmatrix}.
\end{align}
For a circularly polarized incident field, represented by the Jones vector
\begin{align}
\ket{\sigma_\pm}=\frac{1}{\sqrt{2}}
\begin{bmatrix}
1\\
\pm i
\end{bmatrix},
\end{align}
the output field becomes
\begin{align}
J(\theta)\ket{\sigma_\pm}
=
t_+\ket{\sigma_\pm}
+
t_- e^{\mp i2\theta}\ket{\sigma_\mp},
\qquad
t_\pm \equiv \frac{t_{\mathrm{o}}\pm t_{\mathrm{e}}}{2}.
\end{align}
This equation shows that only the cross-circularly polarized component acquires the PB phase $\Phi_{\mathrm{PB}}=\mp 2\theta$, whereas the co-circularly polarized component does not.
Because the co- and cross-circularly polarized components are orthogonal, their intensities are proportional to the squared moduli of the corresponding amplitudes:
\begin{align}
\label{eq:Ico}
I_{\mathrm{co}} &\propto \left|t_+\right|^2
= \frac{1}{4}\left|t_o+t_e\right|^2,\\
\label{eq:Icross}
I_{\mathrm{cross}} &\propto \left|t_-\right|^2
= \frac{1}{4}\left|t_o-t_e\right|^2.
\end{align}
Therefore, the beam-steered component that originates from the cross-circularly polarized field is enhanced when the difference $|t_o-t_e|$ becomes large.

Next, by gradually rotating the Te $c$ axis along a given direction (see e.g. Fig.~\ref{fig:beamsplitter}), a position-dependent PB phase is imposed on the cross-circularly polarized component. For beam steering, we design the spatial distribution of the optic axis orientation $\theta(x)$ as
\begin{align}
\theta(x)=\frac{\pi x}{L},
\end{align}
where $L$ is the spatial period of the metasurface: see Fig.~\ref{fig:sim}(a) for a schematic illustration. In the  experiments, the optic axis orientation is controlled in discrete steps rather than varied continuously. In the present model, however, we assume a continuous rotation of $\theta(x)$ to capture the overall propagation behavior of the deflected beam. Because the optic axis is equivalent under $\theta (x) \rightarrow \theta (x) +\pi$, the distribution induces a PB phase
\begin{align}
    \Phi_{\mathrm{PB}}(x)=\mp 2\theta(x)=\mp \frac{2\pi x}{L},
\end{align}
that varies linearly along $x$ and changes by $2\pi$ over one spatial period $L$. The field immediately after the metasurface is then written as
\begin{align}
\bm{E}_{\mathrm{MS}}(x,y)
=
A(x,y)
\left[
t_+\ket{\sigma_\pm}
+
t_- e^{\mp i2\pi x/L}\ket{\sigma_\mp}
\right],
\end{align}
where $A(x,y)$ denotes the aperture function,  which is introduced to account for the finite lateral size of the metasurface.

Finally, under the Fraunhofer approximation, the far-field distribution is given by the Fourier transform of $\bm{E}_{\mathrm{MS}}(x,y)$:
\begin{align}
\tilde{\bm{E}}_{\mathrm{FF}}(k_x,k_y)
\propto
\iint
\bm{E}_{\mathrm{MS}}(x,y)
e^{-i(k_x x + k_y y)}
\, dx\,dy.
\end{align}
where $k_x$ and $k_y$ are the $x$ and $y$ components of the light wave vector. In evaluating $\tilde{\bm{E}}_{\mathrm{FF}}(k_x,k_y)$, we consider a rectangular metasurface occupying the region of $2 x_0 \times 2 y_0$, where $2x_0$ and $2y_0$ are the lateral dimensions along the $x$ and $y$ directions, respectively. For a rectangular aperture centered at the origin, we write the aperture function as
\begin{align}
A(x,y)=\Pi\!\left(\frac{x}{2x_0}\right)\Pi\!\left(\frac{y}{2y_0}\right),
\end{align}
where $\Pi(u)$ is the rectangular function defined by
\begin{align}
\Pi(u)=
\begin{cases}
1, & |u|<\frac12,\\
0, & |u|>\frac12.
\end{cases}
\end{align}
Using this definition, we obtain
\begin{align}
\label{eq:eff}
\tilde{\bm{E}}_{\mathrm{FF}}(k_x,k_y)
\propto
\ t_+\ket{\sigma_\pm}
\tilde{A}(k_x,k_y)
+
t_-\ket{\sigma_\mp}
\tilde{A}(k_x\mp 2\pi/L,k_y),
\end{align}
where
\begin{align}
\tilde{A}(k_x,k_y)
=
4x_0y_0\,
\mathrm{sinc}(k_x x_0)
\mathrm{sinc}(k_y y_0).
\end{align}
Therefore, the co-circularly polarized component is centered at $k_x=0$, whereas the cross-circularly polarized component is centered at
\begin{align}
k_x=\pm \frac{2\pi}{L}.
\end{align}
Using $k_x=k\sin\beta$ with $k=2\pi/\lambda$ ($\lambda$: the wavelength of the incident light), the deflection angle $\beta$ is given by
\begin{align}\label{eq:defangle}
\sin\beta=\pm \frac{\lambda}{L}.
\end{align}

Note that the same result is obtained from the generalized Snell's law\cite{yu2011science}. For normal incidence, the phase gradient of the PB phase is
\begin{align}
\frac{d\Phi_{\mathrm{PB}}}{dx}=\pm \frac{2\pi}{L},
\end{align}
which leads to
\begin{align}
\sin\beta
=
\frac{\lambda}{2\pi}\frac{d\Phi_{\mathrm{PB}}}{dx}
=
\pm \frac{\lambda}{L}.
\end{align}
Thus, the diffraction picture and the generalized Snell's law give the same steering angle.

\section{Simulations}
\begin{figure}[htbp]
  \centering
  \includegraphics[width=0.98\columnwidth]{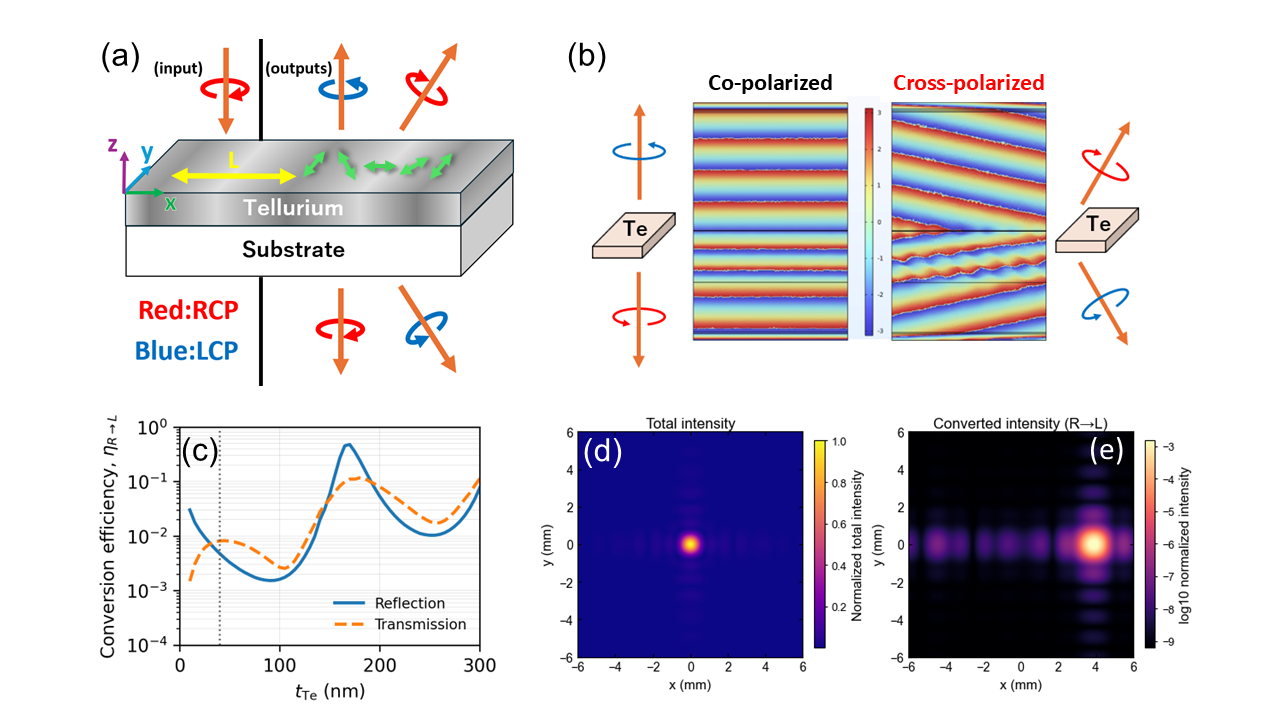}
\caption{\label{fig:sim}
(a) A schematic illustration of the Te beam splitter when right-handed circularly polarized light is incident normally. $L$ is the spatial period of the metasurface. The greyscale contrast in the Te film and the corresponding green double-headed arrows indicate spatial variation of the laser-written $c$-axis orientation (see Fig.~\ref{fig:beamsplitter} caption). The spatial distribution of the $c$ axis orientation enables control of the Pancharatnam-Berry (PB) phase, and thus the beam-splitting functionality.
(b) Calculated reflection and transmission of light from the laser-written Te film. The left (right) panel shows the intensity distribution of the co-circularly polarized (cross-circularly polarized) component of the reflected and transmitted light. The deflected beam is observed only in the right panel. 
(c) Calculated circular-polarization conversion efficiency, $\eta_{\text{R$\to$L}}$, as a function of the Te thickness, $t_{\text{Te}}$, for reflection (blue solid line) and transmission (orange dashed line) geometries. 
The vertical dotted line indicates $t_{\text{Te}} = 40 \ \si{\nano\meter}$, corresponding to the Te thickness used in the experiments. 
(d) Calculated intensity profile of transmitted light from laser-written Te film, plotted in a linear color scale. The displayed intensity is normalized by the maximum total transmitted intensity.
(e) Calculated intensity profile of the cross-circularly polarized component, plotted in a logarithmic color scale. The intensity maximum is shifted to the right (along $+x$), indicating beam deflection caused by the spatially varying optic axis of Te.
} 
\end{figure}

Prior to the experiment, we performed numerical simulations to predict the optical response. Finite-element calculations were carried out using COMSOL Multiphysics. In the simulations, a right-handed circularly polarized (RCP) wave at $\lambda = 1550\ \si{\nano\meter}$ was normally incident from the port at $+z$, while periodic boundary conditions were imposed in the $x$ and $y$ directions to represent the laterally repeating structure. Definition of the coordinate axis and schematic illustration of the sample are presented in Fig.~\ref{fig:sim}(a). The Te layer was treated as an anisotropic medium with $n_o = 4.5$, $n_e = 5.0$, $k_o = 0.05$, and $k_e = 0.03$, experimentally obtained using the method reported in Ref.\cite{mitsuzuka2026photothermaloriginpulselaser}. 
To mimic the experimental condition, the Te layer was divided into six domains whose in-plane optical axes were rotated successively by 30$^\circ$ along the $x$-axis. These six domains constituted one supercell with a period of $L=7.2$ $\si{\micro \meter}$, placed on a substrate with $n = 1.5$ and $k = 0$. From the calculated complex electromagnetic fields, the circular polarization components, $\dfrac1{\sqrt{2}}(E_x \pm i E_y)$, were extracted to examine the phase distributions of the two helicity components. In addition, the power of the transmitted and reflected light were evaluated for each circular polarization component from the corresponding decomposition of the Poynting vector.

Figure~\ref{fig:sim}(a) schematically illustrates the helicity-dependent beam separation produced by the Te beam splitter for the transmitted and reflected light. When a right-handed circularly polarized (RCP) beam is incident normally, the output on each side consists of two circularly polarized components: an undeflected co-circularly polarized component (RCP for the transmitted light and left-handed circularly polarized (LCP) for the reflected light), and a deflected cross-circularly polarized component (LCP for the transmitted light and RCP for the reflected light).
This behavior is confirmed by the simulation results shown in Fig.~\ref{fig:sim}(b). The co-circularly polarized components in the transmitted and reflected light retain an almost uniform phase profile and therefore propagate without deflection. In contrast, the cross-polarized components acquire a spatially varying PB phase, determined by the in-plane orientation distribution of the Te $c$ axis, and are deflected. This behavior is consistent with the theoretical prediction that the PB phase is imparted only to the cross-polarized component. The deflection angle estimated from the simulated field distribution also agrees with the theoretically predicted value. Note that, the sign of the PB phase, and therefore the deflection angle, for the cross-polarized components are reversed when the helicity of the incident light is changed.

We next evaluate the beam-deflection efficiency, where the cross-polarized component corresponds to the deflected beam. The efficiency is defined as the fraction of the cross-polarized component over the total transmitted or reflected power:
\begin{align}\label{eq:conveff}
\eta^\mathrm{t}_{\mathrm{R}\to\mathrm{L}}
=
\frac{P_{\mathrm{L}}^{\mathrm{t}}}{P_{\mathrm{R}}^{\mathrm{t}}+P_{\mathrm{L}}^{\mathrm{t}}},
\qquad
\eta^\mathrm{r}_{\mathrm{R}\to\mathrm{R}}
=
\frac{P_{\mathrm{R}}^{\mathrm{r}}}{P_{\mathrm{R}}^{\mathrm{r}}+P_{\mathrm{L}}^{\mathrm{r}}}.
\end{align}
Here, $P_{\mathrm{R}}^{\mathrm{t}}$ and $P_{\mathrm{L}}^{\mathrm{t}}$ denote the transmitted powers of the right- and left-handed circular polarization components, respectively, while $P_{\mathrm{R}}^{\mathrm{r}}$ and $P_{\mathrm{L}}^{\mathrm{r}}$ denote the corresponding reflected powers. These powers were obtained from the Poynting-vector flux of each circular polarization component.
Figure~\ref{fig:sim}(c) shows the calculated beam-deflection efficiency as a function of Te film thickness. Because of Fabry-P\'{e}rot interference, both the transmittance and the reflectance show complex dependence on the film thickness. For film thicknesses below $300\ \si{\nano\meter}$, the circular-polarization conversion efficiency for a transmission geometry reaches a maximum of approximately $12\%$ at $175\ \si{\nano\meter}$. Experimentally, however, precise laser-induced control of Te $c$ axis orientation has not yet been achieved in films of such thickness. We therefore focus on a film thickness of $40\ \si{\nano\meter}$, which lies within the experimentally accessible range reported previously\cite{kobayashi2026nanolett}. The calculated efficiency for the $40\ \si{\nano\meter}$-thick film is approximately $0.8 \%$. Although the efficiency at this thickness is not particularly high, we considered it sufficient for a proof-of-concept demonstration and therefore performed the experiments at this thickness.

To assess the experimental observability of the deflected beam, we also performed numerical simulations to examine the far-field intensity distribution of the transmitted light for the $40\ \si{\nano\meter}$-thick film. Using the position-dependent Jones matrix method and Fourier optics, we calculated the transmitted field distribution after free-space propagation from the sample. The optical properties of Te were set to be identical to those used in the COMSOL simulations, while the parameters related to the optical setup were chosen to reproduce the experimental configuration. In particular, the incident beam was shaped by a pinhole placed immediately before the metasurface, resulting in a beam radius of $50~\si{\micro\m}$. The simulation results are shown in Fig.~\ref{fig:sim}(d,e). The results indicate that the deflected beam remains spatially distinguishable from the straight-through beam, even though both beams broaden to approximately  $1~\si{\milli \m}$ in radius because of diffraction during propagation. Note that the intensity scale of the plots is adjusted for clarity; the deflected light is much weaker than the straight-through light. Nevertheless, the simulation results suggest that the deflected beam can be observed in the experiment by adjusting the exposure time of the camera to avoid saturation of the straight-through beam while ensuring sufficient contrast for the deflected beam.

\section{Experiments}
\subsection{Sample Fabrication}
\begin{figure}[htbp]
  \centering
  \includegraphics[width=0.8\columnwidth]{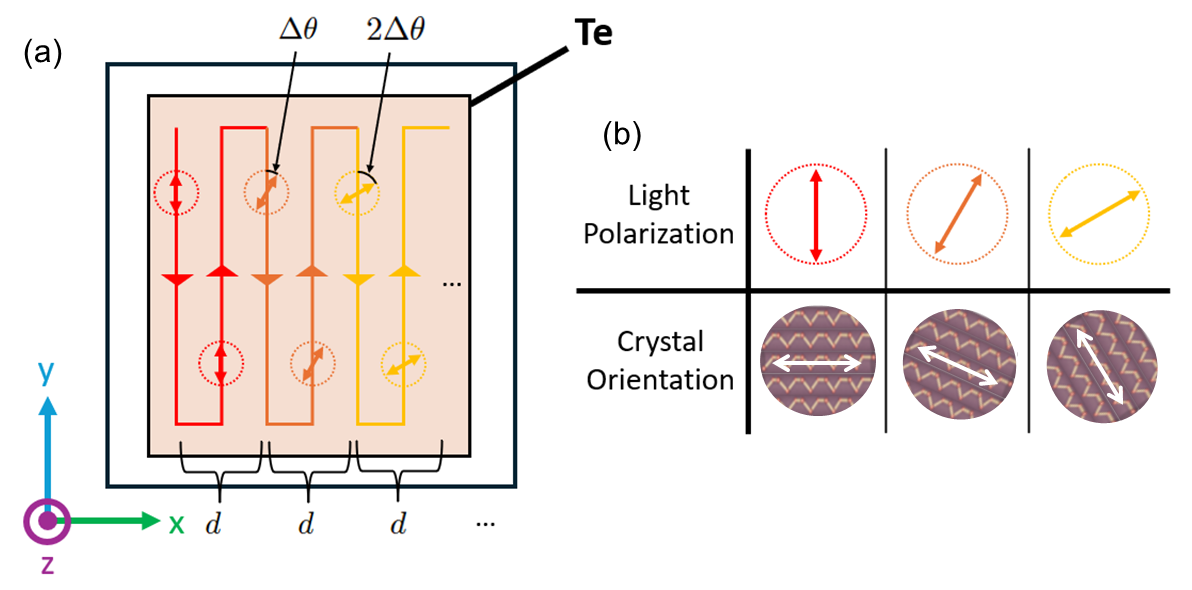}  
\caption{\label{fig:setup} (a,b) Schematic illustration of the pulse laser writing process for fabricating a Te-based beam splitter. A linearly polarized pulse laser is scanned over a polycrystalline Te film using the protocol sketched in (a). 
Since the Te crystal axes are preferentially aligned perpendicular to the incident polarization, the protocol produces a stepwise crystal-axis orientation distribution, as presented in (b). Each pair of adjacent line scan (along $-y$ and $+y$) produces the same crystal-axis orientation. The orientation changes by $\Delta \theta$ between successive scan cycles. 
}
\end{figure}
Films consisting of Te($40 \ \si{\nano \m}$)/Al$_2$O$_3$($2 \ \si{\nano \m}$) were deposited on $c$-Al$_2$O$_3$ substrates using molecular beam epitaxy (MBE). The thickness of the Te layer was confirmed based on X-ray reflectivity (XRR) measurements. The 2 nm-thick Al$_2$O$_3$ layer was deposited as a capping layer to prevent oxidation of the Te layer. The beam deflectors were fabricated by irradiating the Te films at normal incidence with a train of linearly polarized picosecond laser pulses. The laser had a center wavelength of $1030 \ \si{\nano \m}$ and a pulse duration of $4$ - $10 \ \si{\pico \s}$. As depicted in Fig.~\ref{fig:setup}(a), the pulse laser was scanned across a region of the Te film at a scan rate of $0.05 \ \si{\milli \m/ \s}$, corresponding to an estimated pulse number of $N = 2.93 \times 10^6$. The time-averaged power of the pulse laser $P$ was tuned using a neutral density (ND) filter. To investigate the laser-power dependence of the laser-induced reorientation, beam-deflector samples were fabricated using $P = 10, 20, 30, 40, 50, 70,$ and $90 \ \si{\milli\watt}$. The polarization of the laser pulse was tuned using a half-wave plate, and the angle between the $x$-axis and the polarization direction of the laser pulse is denoted by $\theta_{\text{w}}$. $\theta_{\text{w}}$ is changed by $\Delta\theta$ between adjacent regions separated by a distance $d$, that is,
\begin{align}
\label{eq:thetaw}
\theta_{\text{w}}(x,y) = \left[ \frac{x}{d} \right] \Delta \theta.
\end{align}
Note that the Te $c$ axis is oriented perpendicular to the direction specified by $\theta_{\text{w}}$\cite{kobayashi2026nanolett}. The phase profile of the beam deflector is therefore encoded through the stepwise spatial modulation of the writing-polarization angle $\theta_{\text{w}}$. 
The spatial period $L$ of the metasurface is given by $L = \dfrac{\pi}{\Delta \theta} d$. 

\subsection{Measurement setup}
\begin{figure}[htbp]
  \includegraphics[width=0.98\columnwidth]{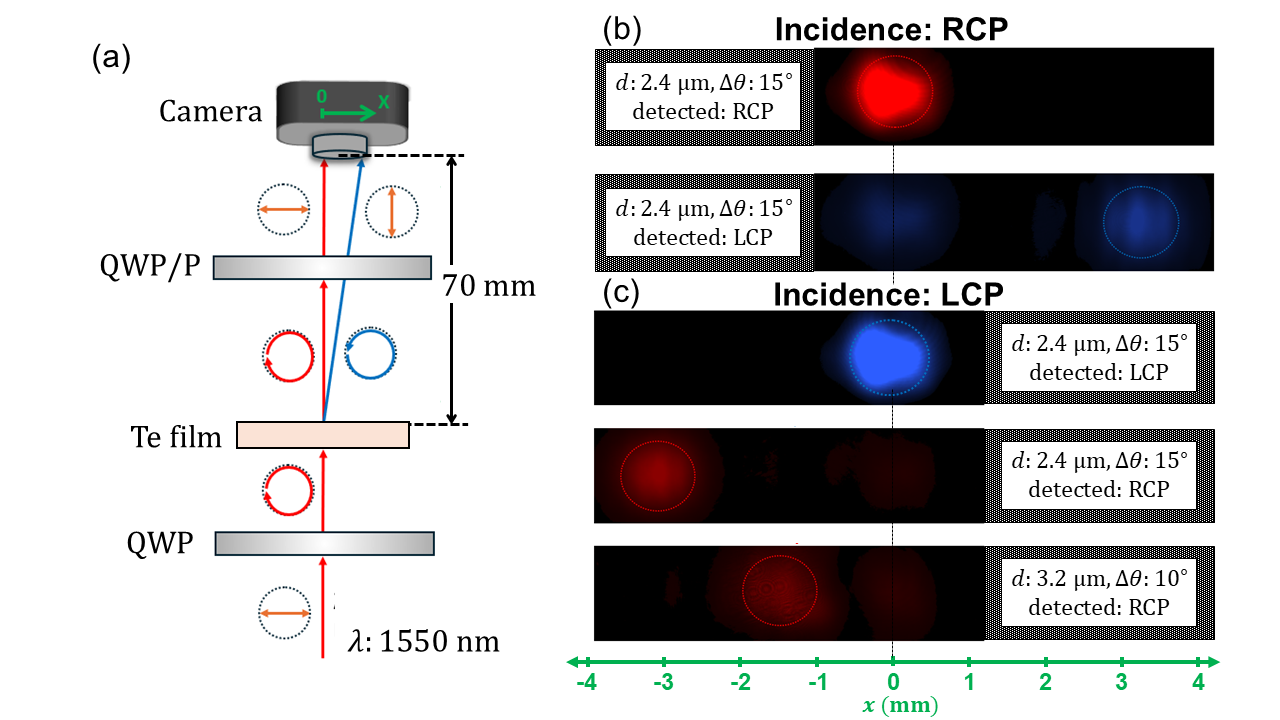}  
  \caption{\label{fig:camera} (a) The setup to measure light deflection. QWP: quarter wave plate, P: polarizer. (b,c) Camera images of the light beam when right(left) handed circularly polarized light is incident on the Te film beam splitter. The $x$-axis shown at the bottom of (c) represents the spatial scale on the camera, and the origin is set at the position where the transmitted beam would arrive in the absence of deflection. The top panel in (b,c) shows an image of the co-polarized beam, whereas the middle and bottom panels display that of the cross-polarized beam. The bottom panel in (c) shows the result for a laser-written Te film with \(d = 3.2~\si{\micro\m}\) and \(\Delta\theta = 10^\circ\), whereas the other panels correspond to a Te film with \(d = 2.4~\si{\micro\m}\) and \(\Delta\theta = 15^\circ\). Here, \(d\) is the width of each domain with the same crystal-axis orientation, and \(\Delta\theta\) is the orientation difference between adjacent domains, as defined in Fig.~\ref{fig:setup}(a). For these parameters, the predicted deflection angle in the bottom panel of (c) is half of the predicted deflection angle in the other panels.}
\end{figure}

To observe the deflection of the light beam by the laser-written Te film, i.e. the beam splitter, we used the experimental setup illustrated in Fig.~\ref{fig:camera}(a). 
A circularly polarized cw laser (wavelength: $1550 \ \si{\nano \m}$) is incident from the normal on the Te beam splitter. 
A pinhole was used to shape the size of the beam similar to that of the beam splitter. The transmitted beam pattern, including both the straight-through and deflected components, was recorded using a infrared camera placed at a distance of $R \simeq 70 \ \si{\milli \m}$ from the film. Here, the straight-through and deflected components correspond to transmitted light with the co- and cross-polarized beams, respectively. The deflection angle $\beta$ was calculated from the position of the light spot in the camera image. The intensities of the straight-through and deflected light were obtained by integrating the pixel values of the corresponding spots in the camera image. 
To compensate for the large power difference between co-polarized and cross-polarized components, the exposure time was adjusted so that the recorded images had comparable contrast without saturation. 
The intensity calculated from the image is normalized by the exposure time.

Within a simple PB phase description, the spatially varying $c$ axis orientation of the Te film is expected to impose a helicity-dependent phase shift on the cross-polarized component. The deflection angle is therefore expected to be proportional to the gradient of the change in the $c$ axis orientation.
The intensity of the deflected light, in contrast, is determined by how efficiently the incident circular polarization is converted into the opposite circular-polarization component in the birefringent Te film. This conversion efficiency is governed by the difference between the ordinary and extraordinary transmission coefficients, $t_\text{o}$ and $t_\text{e}$, as expressed in Eq.~(\ref{eq:Icross}). Accordingly, within this simple theoretical picture, the efficiency of the deflected-light intensity relative to the transmitted-light intensity is given by 
\begin{align}\label{eq:conveff2}
  \text{Efficiency} = \frac{\text{Deflected light intensity}}{\text{Transmitted light intensity}} \propto \frac{|t_\text{o}-t_\text{e}|^2}{|t_\text{o}+t_\text{e}|^2+|t_\text{o}-t_\text{e}|^2}. 
\end{align}

\subsection{Beam deflection imaging}
The images of the transmitted beams captured using the infrared camera are shown in Fig.~\ref{fig:camera}(b,c). Whereas the co-polarized component remains undeflected, the cross-polarized component is deflected. A comparison of these images reveals that the deflection direction reverses when the handedness of the circular polarization of the incident light is reversed, in agreement with the simulation results. 

Samples with different spatial change in the Te $c$ axis orientation, i.e. different pattern of $\theta_\mathrm{w} (x,y)$ [see Eq.~(\ref{eq:thetaw})], exhibit different deflection angles: compare the bottom two images of Fig.~\ref{fig:camera}(c). 
The estimated deflection angle is given by 
\begin{equation}\label{eq:defangle2}
  \beta = \arctan\left(\frac{\lambda}{L}\right) = \arctan\left(\frac{\lambda \Delta\theta}{\pi d}\right).
\end{equation}
Figure~\ref{fig:efficiency}(a,b) compares the estimated deflection angles calculated from Eq.~\eqref{eq:defangle2} with the measured angles. The measured angle is approximately proportional to the estimated value, indicating that the deflection behavior is governed by the designed repetition period. 
The small discrepancy between the measured and estimated angles may be attributed to systematic errors arising from misalignment between the sample and the camera. It may also originate from the discrete nature of the designed phase-shift distribution, i.e. the change in the Te $c$ axis orientation, which leads to deviation from the ideal deflection angle. 

\subsection{Beam splitter conversion efficiency}
\begin{figure}[htbp]
  \includegraphics[width=0.95\columnwidth]{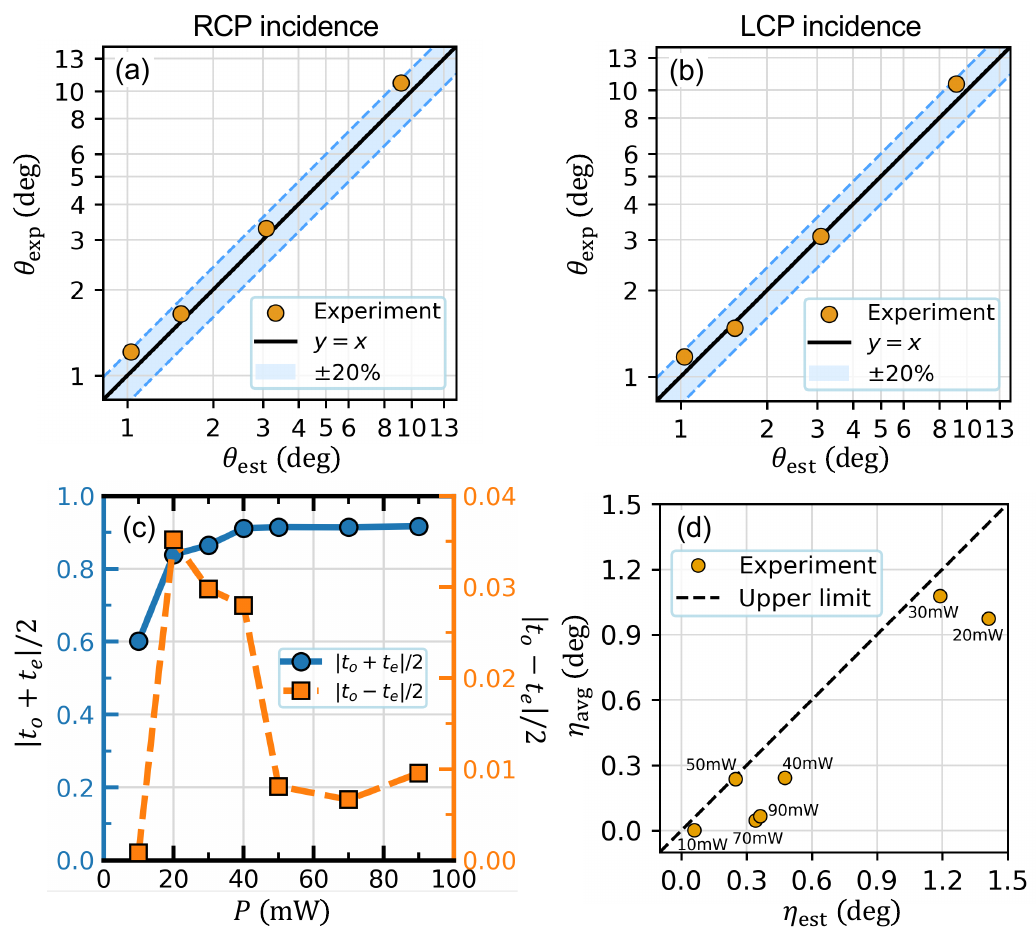}  
  \caption{\label{fig:efficiency} (a,b) Comparison between the measured $\theta_\mathrm{exp}$ and estimated $\theta_\mathrm{est}$ beam-deflection angles for Te film beam splitters under right (a) and left (b) circularly polarized illumination. The horizontal axis shows the deflection angle estimated from the designed phase gradient [Eq.~\eqref{eq:defangle2}], whereas the vertical axis shows the experimentally measured deflection angle. The solid line represents perfect agreement between experimental and estimated values, $\theta_{\text{exp}} = \theta_{\text{est}}$, and the shaded region indicates a $\pm 20\%$ deviation from this relation. (c) Average ($|t_{\mathrm{o}}+t_{\mathrm{e}}|/2$) and difference ($|t_{\mathrm{o}}-t_{\mathrm{e}}|/2$) of the transmission amplitudes of laser-written Te films plotted as a function of time-averaged pulse laser power ($P_\mathrm{laser}$).  
 (d) Comparison between the measured $\eta_\mathrm{avg}$ and estimated $\eta_\mathrm{est}$ efficiency. 
 The dashed line represents the theoretical upper limit, $\eta_{\text{avg}} = \eta_{\text{est}}$. The labels indicate the incident laser power $P_\mathrm{laser}$ used for each measurement.
  }
\end{figure}
Finally, we study the conversion efficiency of the Te beam splitter using laser-written films with different laser power. 
We first measured the polarization-dependent transmission of the irradiated regions to investigate how pulse laser irradiation modifies the optical response, i.e. $|t_e - t_o|/2$ of the laser-written areas of Te. See Supplementary Material, Fig.~\ref{fig:setup:optics}, for the details of the measurement setup. A linearly polarized cw laser with a wavelength of $1550 \ \si{\nano\meter}$ was incident normal to the Te film. The transmitted light was passed through a rotatable linear polarizer, and its intensity was recorded as a function of the rotation angle $\phi$ of the polarizer. 
The angular dependence of the transmitted light intensity was analyzed using a Jones-matrix model. In this model, the irradiated Te film was treated as an anisotropic optical layer with two principal transmission coefficients, $t_o$ and $t_e$, corresponding to the ordinary and extraordinary optical axes, respectively. The transmitted intensity after the polarizer is expressed as
\begin{align}
I(\phi)
=
I_{\mathrm{bg}}
+
I_{\mathrm{in}}
\left|
\bra{p(\phi)}
J(\theta_{\mathrm{Te}})
\ket{p(0)}
\right|^2 ,
\end{align}
where
\begin{align}
J(\theta_{\mathrm{Te}})=
R(-\theta_{\mathrm{Te}})
\begin{pmatrix}
t_o & 0 \\
0 & t_e
\end{pmatrix}
R(\theta_{\mathrm{Te}}), \ \ \ket{p(\phi)} = \begin{pmatrix}
\sin\phi \\ \cos\phi
\end{pmatrix}.
\end{align}
Here, $\theta_{\mathrm{Te}}$ denotes the in-plane orientation angle of the principal optical axes of the irradiated Te film with respect to the lab $x$-axis. $\theta_{\mathrm{Te}}$ is related to the writing-laser polarization angle as $\theta_{\mathrm{Te}} \simeq \theta_{\mathrm{w}}+\pi/2$ since the Te $c$ axis is preferentially aligned perpendicular to the writing polarization\cite{kobayashi2026nanolett}. 
The measured angular dependence of $I(\phi)$ was fitted using this model to estimate $t_o$ and $t_e$ for each irradiated region. 

In Fig.~\ref{fig:efficiency}(c), we plot $|t_{\mathrm{o}}+t_{\mathrm{e}}|/2$ and $|t_{\mathrm{o}}-t_{\mathrm{e}}|/2$ as a function of the writing-laser power $P$. 
$|t_{\mathrm{o}}+t_{\mathrm{e}}|/2$, which characterize the polarization-preserving component, increases with $P$ and approaches a nearly saturated value of $ 1.9$ for $P \gtrsim 50\mathrm{mW}$. In contrast, $|t_{\mathrm{o}}-t_{\mathrm{e}}|/2$, which represents the polarization-converting transmission component, reaches a maximum around $P=20-30~\mathrm{mW}$ and subsequently decreases at higher powers. 
Such reduction in $|t_{\mathrm{o}}-t_{\mathrm{e}}|/2$ with increasing $P_\mathrm{laser}$ is consistent with previous report\cite{kobayashi2026nanolett}. 
Because the deflection of the beam splitter is produced by the cross-polarized component, we estimate the conversion efficiency, $\eta_{\mathrm{est}}$, by substituting $|t_{\mathrm{o}}+t_{\mathrm{e}}|/2$ and  $|t_{\mathrm{o}}-t_{\mathrm{e}}|$ to Eq.~\eqref{eq:conveff2}. This value provides an upper limit for the deflected-light efficiency.

We compare the measured average deflected-light efficiency, $\eta_{\mathrm{avg}}$, with the estimated upper limit.
The measured efficiency is defined as the average of the efficiencies for left- and right-circularly polarized incident light, $\eta_{\text{avg}} = (\eta_{\text{LCP}} + \eta_{\text{RCP}})/2$. 
The experimental results are plotted against $\eta_{\mathrm{avg}}$ in Fig.~\ref{fig:efficiency}(d). The dashed line represents $\eta_{\mathrm{avg}}=\eta_{\mathrm{est}}$, corresponding to the ideal upper limit in which the cross-polarized component is completely directed into the designed diffraction order. The experimental values are generally located below this line, as expected. The deviation of the experimental points from the ideal line indicates that the cross-polarized light is not fully concentrated into the designed deflected order, owing to nonideal factors such as spatial inhomogeneity of the laser-written structure, imperfect phase modulation, and scattering losses. Nevertheless, the measured efficiencies are of the same order as the estimated values and show a similar trend: the largest $\eta_{\mathrm{avg}}$ is obtained for $P=20-30$ mW, where $|t_{\mathrm{o}}-t_{\mathrm{e}}|$ takes a maximum. In the present study, the maximum efficiency of the beam splitter was approximately 1.4\%. As shown in Fig.~\ref{fig:sim}(c), increasing the Te film thickness could substantially improve the conversion efficiency. To achieve such improvement experimentally, laser-based techniques for controlling the $c$ axis orientation in thicker Te films are required. This will be an important direction for developing higher-efficiency Te-based metasurfaces.

\section{Conclusion}
In summary, we demonstrated a Te-based metasurface fabricated by linearly polarized pulse laser irradiation and investigated its optical performance. As a proof-of-concept device, we fabricated a beam splitter and experimentally confirmed helicity-dependent light deflection in the designed direction. The measured deflection angle and efficiency were consistent with the simulated values, indicating that the designed phase distribution was successfully encoded in the Te film. We also found that the deflection efficiency is closely related to the anisotropic transmission response of the laser-written film, which can be tuned by the writing laser power.
These results suggest that the optical response of the device can be improved by controlling the laser-induced anisotropy of the Te film. The findings highlight the potential of laser-written Te as a platform for metasurfaces that can be fabricated without relying solely on lithographic nanopatterning. Because the optical functionality is determined by the written phase distribution, this approach should be extendable to a wider range of metasurface devices based on spatially designed anisotropic optical responses. As the crystal-axis distribution and the anisotropic optical responses can be overwritten repeatedly, the present approach may provide a route toward reconfigurable and/or active metasurfaces based on laser-written anisotropy.

\section*{Acknowledgments}
This work was partly supported by JSPS KAKENHI (Grant Numbers 23H00176,25K22216), MEXT Q-LEAP (Grant Number JPMXS0118067246), MEXT Initiative to Establish Next-generation Novel Integrated Circuits Centers (X-NICS) and Cooperative Research Project Program of RIEC, Tohoku University. T.H. is supported by the Forefront Physics and Mathematics Program to Drive Transformation (FoPM), The University of Tokyo. A.M. is supported by Materials Education program for the future leaders in Research, Industry, and Technology (MERIT), The University of Tokyo. 

\section{Appendix}
\subsection{Finite element method}
\subsubsection{Beam-deflection efficiency}
The finite-element simulations were performed using COMSOL Multiphysics with the Wave Optics Module (Electromagnetic Waves, Frequency Domain interface). The response from a single supercell of the Te beam splitter was calculated. Periodic boundary conditions were applied along the in-plane direction of the Te layer. The incident and transmitted sides were terminated by periodic ports to suppress artificial reflections. The incident field was introduced as a normally incident plane wave, and the transmitted and reflected fields were evaluated on planes sufficiently far from the Te layer.

The complex electric-field distribution obtained from the simulation was decomposed into circular-polarization components using the same convention as in the main text,
\begin{align}
E^{\mathrm{t}}_{\mathrm{R}} &= E^{\mathrm{t}}_x + iE^{\mathrm{t}}_y, 
\qquad
E^{\mathrm{t}}_{\mathrm{L}} = E^{\mathrm{t}}_x - iE^{\mathrm{t}}_y, \\
\qquad
E^{\mathrm{r}}_{\mathrm{R}} &= E^{\mathrm{r}}_x - iE^{\mathrm{r}}_y, 
\qquad
E^{\mathrm{r}}_{\mathrm{L}} = E^{\mathrm{r}}_x + iE^{\mathrm{r}}_y.
\end{align}
This decomposition was used to examine the spatial phase distributions of the two helicity components for the fields transmitted through and reflected from the Te metasurface.
To evaluate the helicity-resolved transmitted and reflected powers, the electromagnetic fields on the evaluation planes were decomposed into the same circular-polarization basis. The power carried by each component was calculated from the corresponding time-averaged Poynting-vector flux through the evaluation plane. Specifically, the transmitted and reflected powers were evaluated as
\begin{align}
P_{\mathrm{R}}^{\mathrm{t}}&=
\int_{A_{\mathrm{t}}}
\left\langle \mathbf{S}_{\mathrm{R}}^{\mathrm{t}} 
\right\rangle
\cdot \hat{\mathbf{n}}_{\mathrm{t}} dA, 
\qquad
P_{\mathrm{L}}^{\mathrm{t}} = \int_{A_{\mathrm{t}}}
\left\langle \mathbf{S}_{\mathrm{L}}^{\mathrm{t}}\right\rangle
\cdot \hat{\mathbf{n}}_{\mathrm{t}} dA\\
P_{\mathrm{R}}^{\mathrm{r}}&=
\int_{A_{\mathrm{r}}}
\left\langle \mathbf{S}_{\mathrm{R}}^{\mathrm{r}} \right\rangle
\cdot \hat{\mathbf{n}}_{\mathrm{r}} dA,
\qquad
P_{\mathrm{L}}^{\mathrm{r}}=
\int_{A_{\mathrm{r}}}
\left\langle \mathbf{S}_{\mathrm{L}}^{\mathrm{r}} \right\rangle
\cdot \hat{\mathbf{n}}_{\mathrm{r}} dA 
\end{align}
where $A_{\mathrm{t}}$ and $A_{\mathrm{r}}$ denote the evaluation planes on the transmission and reflection sides, respectively, and $\hat{\mathbf{n}}_{\mathrm{t}}$ and $\hat{\mathbf{n}}_{\mathrm{r}}$ are the corresponding outward normal vectors. Here, $\left\langle \mathbf{S}_{\mathrm{R}/\mathrm{L}}^{\mathrm{t}} \right\rangle$ and $\left\langle \mathbf{S}_{\mathrm{R}/\mathrm{L}}^{\mathrm{r}} \right\rangle$ denote the time-averaged Poynting vectors associated with the $E_{\mathrm{R}/\mathrm{L}}$ component of the transmitted and reflected fields, respectively. By substituting these values into Eq.~\eqref{eq:conveff}, we calculated the beam-deflection efficiencies and examined their dependence on the Te film thickness, as shown in Fig.~\ref{fig:sim}(c).

\subsubsection{Intensity profile}
Here we describe how we obtained the transmitted light intensity profile presented in Fig.~\ref{fig:sim}(d) and (e).
The calculations were performed using the angular spectrum method in Python. We first defined the field distribution before the metasurface $\bm{E}_\text{in}(x,y)$ as a right-circularly polarized Gaussian mode with a beam spot size of $5~\mathrm{mm}$, truncated by a circular aperture with a diameter of $0.1~\mathrm{mm}$. The field distribution immediately after the metasurface $\bm{E}_\text{MS}(x,y)$ was then calculated by applying the position-dependent Jones matrix to the incident field, as described in the main text. The parameters for the Jones matrix were set based on the thickness of the Te layer and its anisotropic transmission properties. Finally, we computed the field distribution at the observation plane using the angular spectrum method.

The angular spectrum method is a Fourier optics technique that allows us to calculate the propagation of light fields through free space. First, we compute the Fourier transform of the field distribution at the metasurface
\begin{align}
\tilde{\bm{E}}_{\mathrm{MS}}(k_x,k_y) = \iint \bm{E}_{\mathrm{MS}}(x,y) e^{-i(k_x x + k_y y)} \, dx\,dy
\end{align} 
to obtain the angular spectrum. Then, in order to account for the distance between the metasurface and the observation plane, we apply a propagation factor $e^{i k_z z}$ to each component of the angular spectrum
\begin{align}
\tilde{\bm{E}}_{\mathrm{FF}}(k_x,k_y) =  \tilde{\bm{E}}_{\mathrm{MS}}(k_x,k_y) e^{i k_z z},
\end{align}
where $k_z=\sqrt{k^2-k_x^2-k_y^2}$ ($k$ : free-space wavenumber).  Unlike the simplified far-field treatment used in the main text, the present calculation directly propagates the angular spectrum to the observation plane without invoking the Fraunhofer approximation.
Finally, we take the inverse Fourier transform to obtain the field distribution at the observation plane
\begin{align}
\bm{E}_{\mathrm{FF}}(x,y) = \frac{1}{(2\pi)^2}\iint \tilde{\bm{E}}_{\mathrm{FF}}(k_x,k_y) e^{i(k_x x + k_y y)} \, dk_x\,dk_y.
\end{align}
After obtaining the far-field distribution, we calculated the intensity profiles for both the co-polarized and cross-polarized components by taking the squared modulus of the corresponding field components. The resulting intensity distributions were then plotted to generate Fig.~\ref{fig:sim}(d) and (e).

\subsection{Measurement setup}
\begin{figure}[htb]
  \includegraphics[width=0.95\columnwidth]{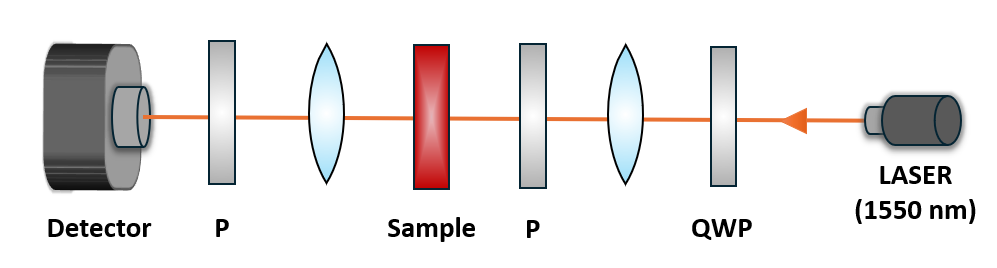}  
  \caption{\label{fig:setup:optics} The optical setup to determine the transmission coefficients $t_o$ and $t_e$ of the laser-written areas of Te film. QWP: quarter wave plate, P: polarizer.}
\end{figure}

In Fig.~\ref{fig:setup:optics}, we present the optical setup used to extract the transmission coefficients $t_o$ and $t_e$, corresponding to those along the ordinary and extraordinary optical axes, respectively. The transmission intensity was measured as a function of the rotation angles of the two polarizers, and the resulting polarization-dependent intensity variation was used to determine the transmission coefficients.

\bibliography{refs_060826}

\end{document}